# A perspective on meta-boundaries

Xinyan Zhang[1,2], Jialin Chen[1,2,3], Ruoxi Chen[1,2], Chan Wang[1,2], Tong Cai[1,4,*], Hongsheng Chen[1,2,*], and Xiao Lin[1,2,*]

[1]Interdisciplinary Center for Quantum Information, State Key Laboratory of Modern Optical Instrumentation, ZJU-Hangzhou Global Scientific and Technological Innovation Center, Zhejiang University, Hangzhou 310027, China.

[2]International Joint Innovation Center, Key Lab. of Advanced Micro/Nano Electronic Devices & Smart Systems of Zhejiang, The Electromagnetics Academy at Zhejiang University, Zhejiang University, Haining 314400, China.

[3]Department of Electrical and Computer Engineering, Technion-Israel Institute of Technology, Haifa 32000, Israel.

[4]Air and Missile Defense College, Air Force Engineering University, Xi'an 710051, China.

[*]Corresponding authors. Email: xiaolinzju@zju.edu.cn (X. Lin); caitong326@zju.edu.cn (T. Cai); hansomchen@zju.edu.cn (H. Chen)

**The judicious design of electromagnetic boundary provides a crucial route to control light-matter interactions, and it is thus fundamental to basic science and practical applications. General design approaches rely on the manipulation of bulk properties of superstrate or substrate and on the modification of boundary geometries. Due to the recent advent of metasurfaces and low-dimensional materials, the boundary can be flexibly featured with a surface conductivity, which can be rather complex but provide an extra degree of freedom to regulate the propagation of light. In this perspective, we denote the boundary with a non-zero surface conductivity as the meta-boundary. The meta-boundaries are categorized into four types, namely isotropic, anisotropic, biisotropic and bianisotropic meta-boundaries, according to the electromagnetic boundary conditions. Accordingly, the latest development for these four kinds of meta-boundaries are reviewed. Finally, an outlook on the research tendency of meta-boundaries is provided, particularly on the manipulation of light-matter interactions by simultaneously exploiting meta-boundaries and metamaterials.**

**Introduction**

When two different bulk media are stacked together, an electromagnetic boundary would be formed. The electromagnetic boundary can be exploited to flexibly control the propagation of light, including its phase, amplitude, polarization, and direction [1-10]. This way, the electromagnetic boundary is crucial to control light-matter interactions, and its continuing exploration has led to many exotic phenomena and practical applications, including negative refraction [11-16], superlens [17-18], cloak [19-24], and DB boundary [25-27].

Generally, the design of electromagnetic boundary relies on changing the geometric shape of the boundary (e.g. grating) or on changing the optical properties of bulk media, which may function as the superstrate or substrate of the boundary. Thanks to the recent development of metamaterials, the optical properties of bulk media can be tailored in a desired manner [28-36]. For instance, by following the design methodology of metamaterials, the effective optical response of bulk media can range from being isotropic, anisotropic, biisotropic to being bianisotropic [37-40] and is closely related to Maxwell equations, which are the basis to describe the light-matter interaction. The Maxwell equations are

$$\nabla \times \bar{H} = \frac{\partial}{\partial t}\bar{D} + \bar{J} \tag{1}$$

$$\nabla \times \bar{E} = -\frac{\partial}{\partial t}\bar{B} \tag{2}$$

$$\nabla \cdot \bar{D} = \rho \tag{3}$$

$$\nabla \cdot \bar{B} = 0 \tag{4}$$

where $\bar{J}$ and $\rho$ represent the electric current density and electric charge density, respectively; the relations between the electric field $\bar{E}$, magnetic field $\bar{H}$, electric displacement $\bar{D}$, and magnetic flux density are described by the constitutive relations. That is, the constitutive relations serve as the complementary but important information for Maxwell equations and can well describe the optical response of bulk media. To be specific, the constitutive relations for bulk isotropic media [41] can be readily expressed as

$$\bar{D} = \varepsilon \cdot \bar{E} \tag{5}$$

$$\bar{B} = \mu \cdot \bar{H} \tag{6}$$

where $\varepsilon$ and $\mu$ represent the permittivity and permeability, respectively. Similarly, the constitutive relations for anisotropic media [41] are written as

$$\bar{D} = \bar{\bar{\varepsilon}} \cdot \bar{E} \tag{7}$$

$$\bar{B} = \bar{\bar{\mu}} \cdot \bar{H} \tag{8}$$

where $\bar{\bar{\varepsilon}}$ and $\bar{\bar{\mu}}$ become to a tensor (namely a $3 \times 3$ matrix). By contrast, the constitutive relations for biisotropic media [41] are

$$\bar{D} = \varepsilon \cdot \bar{E} + \xi \cdot \bar{H} \tag{9}$$

$$\bar{B} = \mu \cdot \bar{H} + \zeta \cdot \bar{E} \tag{10}$$

where $\xi$ and $\zeta$ describe the magnetoelectric coupling. Similarly, the constitutive relations for bianisotropic media [41] can be expressed as

$$\bar{D} = \bar{\bar{\varepsilon}} \cdot \bar{E} + \bar{\bar{\xi}} \cdot \bar{H} \tag{11}$$

$$\bar{B} = \bar{\bar{\mu}} \cdot \bar{H} + \bar{\bar{\zeta}} \cdot \bar{E} \tag{12}$$

where $\bar{\bar{\xi}}$ and $\bar{\bar{\zeta}}$ become to a tensor. The diversity of bulk media as governed by equations (5-12) indicates the enormous possibilities to conceive different types of electromagnetic boundary.

Apart from changing the optical properties of bulk media, the electromagnetic boundary itself can be directly tailored, due to the recent advent of metasurfaces [42-46] and low-dimensional materials [47-54], which can be well modelled by an effective two-dimensional surface without a thickness but with a non-zero surface conductivity [55-64]. In other words, with the addition of metasurfaces or low-dimensional materials, the boundary can be featured with a non-zero surface conductivity. This way, the judicious design of surface conductivity can provide an extra degree of freedom to regulate the electromagnetic boundary conditions and thus the light-matter interaction. For the simplicity of conceptual illustration in this work, the boundary without surface conductivities is denoted as the common boundary; by contrast, the boundary with non-zero surface conductivities is termed as the meta-boundary; see the schematic in Figure 1a&b.

Due to the abundance of metasurfaces and low-dimensional materials, the electromagnetic boundary conditions can have different mathematical forms. According to these different forms of electromagnetic boundary conditions, the meta-boundary in principle can be categorized into four types, namely isotropic, anisotropic, biisotropic and bianisotropic meta-boundaries; see the summarization of boundary conditions for meta-boundaries in Figure 1c. Such categorization for meta-boundaries based on the boundary conditions is intrinsically analogous to the studies for bulk media, whose categorization is based on the constitutive relations.

To be specific, the boundary conditions for isotropic meta-boundary are

$$\hat{n} \times (\bar{E}_1 - \bar{E}_2) = \sigma_m \cdot (\bar{H}_1 + \bar{H}_2)/2 \tag{13}$$

$$\hat{n} \times (\bar{H}_1 - \bar{H}_2) = \sigma_e \cdot (\bar{E}_1 + \bar{E}_2)/2 \tag{14}$$

where $\sigma_e$ and $\sigma_m$ stand for the electric and magnetic surface conductivities, respectively; $\bar{E}_1$ or $\bar{E}_2$ and $\bar{H}_1$ or $\bar{H}_2$ are the electric and magnetic fields in the superstrate (denoted as region 1 in Figure 1a&b) or substrate (region 2) very close to the boundary, respectively; and $\hat{n}$ is the surface normal.

When the meta-boundary becomes to be anisotropic, the boundary conditions in equations (13-14) are changed to

$$\hat{n} \times (\bar{E}_1 - \bar{E}_2) = \bar{\bar{\sigma}}_m \cdot (\bar{H}_1 + \bar{H}_2)/2 \tag{15}$$

$$\hat{n} \times (\bar{H}_1 - \bar{H}_2) = \bar{\bar{\sigma}}_e \cdot (\bar{E}_1 + \bar{E}_2)/2 \tag{16}$$

where $\bar{\bar{\sigma}}_e$ and $\bar{\bar{\sigma}}_m$ are a tensor (namely a $2 \times 2$ matrix).

If the meta-boundary is biisotropic, the magnetoelectric coupling would appear in the boundary conditions. Then the boundary conditions for biisotropic meta-boundary become to

$$\hat{n} \times (\bar{E}_1 - \bar{E}_2) = \sigma_m \cdot (\bar{H}_1 + \bar{H}_2)/2 + \sigma_\xi \cdot (\bar{E}_1 + \bar{E}_2)/2 \tag{17}$$

$$\hat{n} \times (\bar{H}_1 - \bar{H}_2) = \sigma_e \cdot (\bar{E}_1 + \bar{E}_2)/2 + \sigma_\zeta \cdot (\bar{H}_1 + \bar{H}_2)/2 \tag{18}$$

where the surface conductivities of $\sigma_\xi$ and $\sigma_\zeta$ represent the magnetoelectric coupling.

If the meta-boundary is bianisotropic, the general form for boundary conditions are

$$\hat{n} \times (\bar{E}_1 - \bar{E}_2) = \bar{\bar{\sigma}}_m \cdot (\bar{H}_1 + \bar{H}_2)/2 + \bar{\bar{\sigma}}_\xi \cdot (\bar{E}_1 + \bar{E}_2)/2 \tag{19}$$

$$\hat{n} \times (\bar{H}_1 - \bar{H}_2) = \bar{\bar{\sigma}}_e \cdot (\bar{E}_1 + \bar{E}_2)/2 + \bar{\bar{\sigma}}_\zeta \cdot (\bar{H}_1 + \bar{H}_2)/2 \tag{20}$$

where $\bar{\bar{\sigma}}_\xi$ and $\bar{\bar{\sigma}}_\zeta$ also become to a tensor.

From equations (13-20), the meta-boundary provides an exotic way to tailor the electromagnetic boundary conditions and is thus of paramount importance for the exploration of novel light-matter interactions. We then briefly review the recent progress in the realm of meta-boundaries, followed by a perspective on their research tendency. Particularly, due to the infinite vitality of metamaterials and meta-boundaries, the combination of metamaterials and meta-boundaries is promising to provide a powerful platform for the arbitrary manipulation of light.

**Isotropic meta-boundary**

As indicated in equations (13-14), the isotropic meta-boundary requires the emergence of isotropic electric and/or magnetic surface conductivities at the boundary. These surface conductivities are typically constructed, for example, by exploiting metal-based metasurfaces [55-64] in Figure 2a-c, all-dielectric metasurfaces [65-68], monolayer graphene [50,69] in Figure 2d, and the inversion layer at the insulator-semiconductor interface (which acts as a two-dimensional electron gas and is similar to the role of graphene) [70]. Moreover, due to the recent advances in nanofabrication, these surface conductivities based either on metasurfaces or low-dimensional materials can be actively tunable [71-78].

The elaborate choice of electric and/or magnetic surface conductivities at the isotropic meta-boundary enables many exotic applications, such as the design of Huygens' surface [79,80], frequency-selective surface [81-83], and high-impedance surface. With the existence of Huygens' surface in Figure 2a, the transmitted light can propagate along a direction, which is not normal to the meta-boundary, under the normal incidence [80]. When considering the frequency dispersion of these surface conductivities, the isotropic meta-boundary can be transparent only to light incidence with certain frequencies and functions as a frequency-selective surface. When integrating the isotropic meta-boundary with input ports, the isotropic holographic metasurfaces for dual-functional radiations without mutual interferences can be obtained [61], as shown in Figure 2b. In addition, while the Brewster effect [41,84-89] for transverse-electric (TE) waves is believed to exist only in systems with magnetic responses, the isotropic meta-boundary with a specific electric surface conductivity can give rise to the TE Brewster effect in a homogeneous dielectric interface without magnetic responses [84].

The isotropic meta-boundary is also widely used in controlling the light flow at the subwavelength scale, especially for the propagation of surface waves. According to the boundary conditions in equations (13-14), the dispersion for transverse-magnetic (TM) or TE surface waves supported at the isotropic meta-boundary can be derived as

$$\text{TM surface waves: } \left(1 + \frac{\sigma_e k_z}{2\omega\varepsilon}\right)\left(1 - \frac{\sigma_m k_z}{2\omega\mu}\right) = 0 \qquad (21)$$

$$\text{TE surface waves: } \left(\frac{k_z}{\omega\mu} + \frac{\sigma_e}{2}\right)\left(\frac{k_z}{\omega\varepsilon} - \frac{\sigma_m}{2}\right) = 0 \qquad (22)$$

For illustration, here we consider the symmetric structure in Figure 1a, namely region 1 and region 2 are the same and have a permittivity $\varepsilon$ and a permeability $\mu$; and $k_z$ stands for the component of wavevector perpendicular to the meta-boundary. If the environment is composed of positive-index materials, namely $\varepsilon > 0$ and $u > 0$, the emergence of TM surface waves in isotropic meta-boundaries requires $\text{Im}(\sigma_e) > 0$ and/or $\text{Im}(\sigma_m) < 0$ according to equation (21); by contrast, the existence condition for TE surface waves in isotropic meta-boundaries becomes to $\text{Im}(\sigma_e) < 0$ and/or $\text{Im}(\sigma_m) > 0$ according to equation (22) [49].

If the isotropic meta-boundary is composed of the monolayer graphene, the spatial confinement of TM graphene plasmons is generally much better than that of TE graphene plasmons [90-93]. For example, the wavelength of TM graphene plasmons (Figure 2d) can be two orders of magnitude smaller than the wavelength of light in free space, while the wavelength of TE graphene plasmons is similar to the wavelength of light in free space. Moreover, due to the low material loss and active tunability of graphene, there are extensive studies about TM surface waves in isotropic meta-boundaries with graphene. On the other hand, how to improve the spatial confinement of TE surface waves in isotropic meta-boundaries with graphene, especially in terahertz or infrared regimes, remains a challenging issue. One way to tackle this issue is to replace the positive-index environment of the isotropic meta-boundary with the negative-index environment (which has $\varepsilon < 0$ and $u < 0$) [94]. Under this scenario, it is worthy to note that the existence conditions for both TM and TE surface waves in isotropic meta-boundaries would be drastically changed, according to equations (13-14); see systematic discussions in Ref. [94].

**<u>Anisotropic meta-boundary</u>**

One typical feature of anisotropic meta-boundaries is their distinct response to TE and TM waves.

That is, the anisotropic meta-boundary is sensitive to the incident polarization of light and can be exploited to perform many polarization-based functions, including polarization selectivity, polarization conversion, and perfect absorption of light with specific absorption [68,95-102]. For example, the application of anisotropic meta-boundaries can result in the spatial separation of light with different linear polarizations [97]. Utilizing single or multiple vertically-parallel anisotropic meta-boundaries can realize the conversion between the circularly-polarized light (Figure 3a) and linearly-polarized light [103]. If the anisotropic meta-boundary has certain nonlinear effect, the laser-mode locking can be achieved; that is, only the laser mode with a specific polarization at the prescribed frequency can be retained in Figure 3b [104]. Figure 3c shows that the anisotropic meta-boundary plays an important role in spatially molding the propagation of surface waves [105].

Among various anisotropic meta-boundaries, the hyperbolic meta-boundary is of particular interest, whose corresponding electric and/or magnetic surface conductivities have $\sigma_{e,xx} \cdot \sigma_{e,yy} < 0$ and/or $\sigma_{m,xx} \cdot \sigma_{m,yy} < 0$, where $\bar{\bar{\sigma}}_e = \text{diag}[\sigma_{e,xx}\ \sigma_{e,yy}]$ and $\bar{\bar{\sigma}}_m = \text{diag}[\sigma_{m,xx}\ \sigma_{m,yy}]$. One key advantage of hyperbolic meta-boundaries is their capability to support hyperbolic plasmons (see their near-field excitation in Figure 3d for example), whose iso-frequency contour is hyperbolic [105-111]. Due to the highly squeezed nature of hyperbolic plasmons in space, hyperbolic meta-boundaries have been extensively used to manipulate the flow of nano-light, especially for the realization of propagation with high directionality. Therefore, the hyperbolic meta-boundaries enable a variety of optical functions at the deep-subwavelength scale, such as plasmonic super-lens, enhanced spontaneous emission, and directional guidance at the nanoscale [112-120]. On the other hand, low-dimensional materials provide an abundant choice for the construction of hyperbolic meta-boundaries [121-124]. For example, hyperbolic meta-boundaries can be fabricated by patterning isotropic low-dimensional materials (e.g. nanoribbons arrays of graphene or hexagonal boron nitride (BN) in Figure 3d-e or by using naturally anisotropic materials (e.g. doped monolayer black phosphorus in Figure 3f) [47,109,125].

**<u>Biisotropic meta-boundary</u>**

Distinct from the isotropic and anisotropic meta-boundaries, the biisotropic meta-boundary has the magnetoelectric coupling [126-135], which is modelled by the magnetoelectric surface conductivities in equations (17, 18). In order to introduce the magnetoelectric coupling, the electromagnetic boundary becomes relatively complex in the structural fabrication and should be carefully designed. The realization of biisotropic meta-boundaries may require, for example, the usage of single metasurface with some irregular bulges in Figure 4a or the vertical stacking of multiple metasurfaces with a special interlayer twisted angle in Figure 4b [129,130]. Moreover, to ensure the isotropy of magnetoelectric coupling, these metasurfaces should have a good rotational symmetry. Figure 3c shows that the time-modulated metasurface, which consists of metallic patches array and parallel capacitor, may provide another route for the design of biisotropic meta-boundary [131]. Meanwhile, the biisotropic boundary can be achieved by inserting chiral materials (which intrinsically have the isotropic magnetoelectric coupling) into an ultrathin dielectric slab in Figure 3d [132]. With the help of the biisotropic meta-boundary, the scattering cross section of certain objects can be reduced significantly [134]. If the biisotropic meta-boundary is spatio-temporally modulated, one may further realize the nonreciprocal transmission of light [131].

Despite these recent progresses, the fabrication of biisotropic meta-boundaries still lacks a systematic methodology and remains a challenge. Accordingly, the biisotropic meta-boundary has been rarely explored, when compared with the isotropic and anisotropic ones.

**Bianisotropic meta-boundary**

For bianisotropic meta-boundaries, their fabrication [136-138] can be relatively easier when compared with biisotropic meta-boundariers, since there is no need to keep the isotropy of magnetoelectric coupling. For instance, the metasurface with the Omega type meta-atom, which has strong anisotropic magnetoelectric coupling and is relatively simple in fabrication in Figure 5a [139,140], provides a typical route to construct the bianisotropic meta-boundary. Apart for metal-based metasurfaces, the bianisotropic meta-boundary can also be designed by arrays of all-dielectric cylindrical rods with an irregular hole. Figure 5b shows that this kind of bianisotropic meta-boundary can have a large nonlinear effect, which can induce a giant asymmetric second harmonic generation [136]. Moreover, recent works show that the twisted bilayer graphene in Figure 5c can facilitate the design of novel bianisotropic meta-boundaries, since the twisted bilayer graphene intrinsically possesses the bianisotropy, which originates from the interlayer quantum coupling [141-147]. Moreover, the bianisotropic meta-boundary assisted by the twisted bilayer graphene can support the propagation of chiral plasmons, which have not only the transverse spin but also the longitudinal spin [144]. This way, new kinds of spin-orbit interaction of light are expected in these bianisotropic meta-boundaries but await further exploration both in theory and experiments.

Despite the complexity of their electromagnetic boundary conditions in equations (19,20), bianisotropic meta-boundaries have demonstrated unique applications in the manipulation of light-matter interactions. For example, bianisotropic meta-boundaries is capable to realize the complete polarization conversion during either the reflection or transmission process in Figure 5d [143,148]. With the help of bianisotropic meta-boundaries, the polarization of free-electron radiation (e.g. Smith-Purcell radiation in Figure 5e) can be arbitrarily designed [46,149]. In addition, the bianisotropic meta-boundary is widely used in other realms, including the transformation of propagating waves into surface waves, self-isolated Raman lasing in Figure 5f, and asymmetric transmission of light [150, 151].

**<u>Active meta-boundary</u>**

One rising tendency for the further exploration of meta-boundary is the investigation of active (or temporal) meta-boundary, which is temporally modulated and thus whose corresponding surface conductivity is a function of time. Due to the existence of time modulation, the active meta-boundaries can be exploited to realize exotic performance, including the dynamical beam steering, non-reciprocal transmission of light, real-time on-chip communications, and doppler-like frequency shift of light [152-161]. Figure 6a shows that the active meta-boundary can tune the frequency of reflected light and converge the reflected light into any predesigned focal point [158].

To further improve the capability of meta-boundaries, one may add the spatial modulation into the active meta-boundary. Such kinds of active (or spatio-temporal) meta-boundaries can be achieved by actively and separately tuning each unit cell of metasurface, through the electronic or optical programming [162-167]. These active meta-boundaries can perform many advanced functions, such as intelligent communication and holographic imaging. Figure 6b shows one typical active meta-boundary, which is created by arrays of complementary metal-oxide-semiconductor (CMOS) - based chip tiles and can digitally control the amplitude and phase of light [163]. These active meta-boundaries are then advantageous in controlling the scattering of light in real time. For example, Figure 6c shows that the real-time self-adaptive cloak can be implemented based on the active meta-boundary with the aid of artificial intelligence [168]. To be specific, though inserting an artificial neural network between the detection system and the control system, the active meta-boundary behaves as a carpet cloak with a very small scattering cross section, regardless of the real-time change of surrounding environment.

**<u>Multiple meta-boundaries</u>**

Apart from individual meta-boundary, multiple meta-boundaries can be combined, which may induce the in-plane or out-of-plane interactions between neighboring meta-boundaries. Due to the rich yet relatively-less explored physics in these in-plane and out-of-plane interactions, the exploration of multiple meta-boundaries is becoming another research tendency [169-175].

To induce the in-plane interaction, one way is to splice two different meta-boundaries in a same plane, as schematically shown in Figure 7a. As a result, the in-plane splicing between two meta-boundaries can give rise to a high-order boundary, namely a one-dimensional line interface. Correspondingly, the existence of high-order boundary can induce the phenomena of reflection, transmission and even scattering for surface waves. For example, Figure 7b shows the in-plane splicing between two hyperbolic meta-boundaries, which are the same but have different orientation angles. Remarkably, the appearance of high-order boundary between these two hyperbolic meta-boundaries can support the phenomenon of all-angle negative refraction of highly squeezed hyperbolic plasmons within a broad frequency regime [176]. On the other hand, it is worthy to note that during the reflection and transmission process of surface waves, the scattering of surface waves into propagating waves generally exists, which indicates an unwanted degradation of signals and may induce a noisy electromagnetic background. Therefore, the suppression of this scattering has been long sought after but is still challenging in experiments. The realization of this enticing goal based on the high-order boundary in Fig. 7a, which is formed by diverse meta-boundaries, might be promising but remains un-explored.

To induce the out-of-plane interaction between meta-boundaries, one way is to vertically stack multiple parallel but spatially-separated meta-boundaries, as shown in Figure 7c. For example, Figure 7d shows the schematic structural of double bilayer graphene, separated by thin BN slabs. Remarkably, this kind of multiple meta-boundaries can support the emergence of excitonic superfluid phase [177]. Moreover, the out-of-plane interaction between neighboring meta-boundaries would become more complex, if there exists the interlayer twist angle [178-188], as schematically shown in Figure 7e. For example, Figure 7f shows the multiple meta-boundaries composed of rotated multiple bilayer graphene [189]. The canalized excitation and propagation of surface waves can be achieved by stacking multiple $\alpha$-phase molybdenum trioxide ($\alpha$-MoO$_3$) slabs, in which an ultrathin slab of $\alpha$-MoO$_3$ may be approximately modelled by an anisotropic metasurface [179]. Moreover, the phenomenon of topological transition for the iso-frequency contour of surface waves would occur, readily by rotating the interlayer twist angle.

**Composite structures composed of meta-boundaries and metamaterials**

Meta-boundaries have shown impressive capabilities to tailor the optical response of the interface, while metamaterials have shown impressive capabilities to tailor the optical response of bulk media. Therefore, it is straightforward to combine meta-boundaries and metamaterials in a same structure (see the schematic in Figure 8a). Since both meta-boundaries and metamaterials can be isotropic, anisotropic, biisotropic and bianisotropic, there are various possible combinations between them [190-194]; see the brief schematic summarization in Figure 8b. Then the composite structures, which simultaneously have meta-boundaries and metamaterials, are promising to provide a powerful yet plentiful platform for the manipulation of light-matter interactions. Therefore, these composite structures are worthy more in-depth and systematic exploration

For example, Figure 9a-b show two composite structures by depositing a two-dimensional transition metal dichalcogenides (TMD) material on a grating or a two-dimensional photonic crystal [190,191], both of which can be treated as a metamaterial with strong nonlocality. These composite structures can flexibly tailor the topological property of polaritonic systems. Figure 9c shows another composite structure by placing a monolayer graphene on a uniaxial BN slab with a finite thickness. This composite structure is able to realize the all-angle negative refraction of highly squeezed isotropic surface waves [11]. Figure 9d further shows that the composite structure composed of the monolayer graphene and an $\alpha$-$MoO_3$ slab can facilitate the realization of polaritonic focus [186].

In conclusion, we have highlighted the importance of meta-boundaries in controlling the light-matter interactions. After reviewing the recent progresses on isotropic, anisotropic, biiostropic and bianisotropic meta-boundaries, the research tendencies for meta-boundaries are analyzed, including the continuing exploration of active meta-boundaries, multiple meta-boundaries, and composite structures composed of meta-boundaries and metamaterials. Regarding composite structures, most current researches are mainly carried out for relatively-simple composite structures, which are composed of isotropic or anisotropic meta-boundaries and isotropic or anisotropic metamaterials. The relatively-complex composite structure, such as those composed of bianisotropic meta-boundaries and bianisotropic metamaterials, demands more extensive and in-depth studies.

**ACKNOWLEDGMENTS**
The work at Zhejiang University was sponsored in part by the National Natural Science Foundation of China (NSFC) under Grants No. 62175212, the National Natural Science Fund for Excellent Young Scientists Fund Program (Overseas) of China, the Fundamental Research Funds for the Central Universities (2021FZZX001-19), Zhejiang University Global Partnership Fund, the Key Research and Development Program of the Ministry of Science and Technology under Grants No. 2022YFA1404704, 2022YFA1404902, and SQ2022YFA1400025, the National Natural Science Foundation of China (NNSFC) under Grants No.11961141010 and No. 61975176, the Fundamental Research Funds for the Central Universities, and the Chinese Scholarship Council (CSC No. 202206320287).

**NOTES**

The authors declare no competing financial interest.

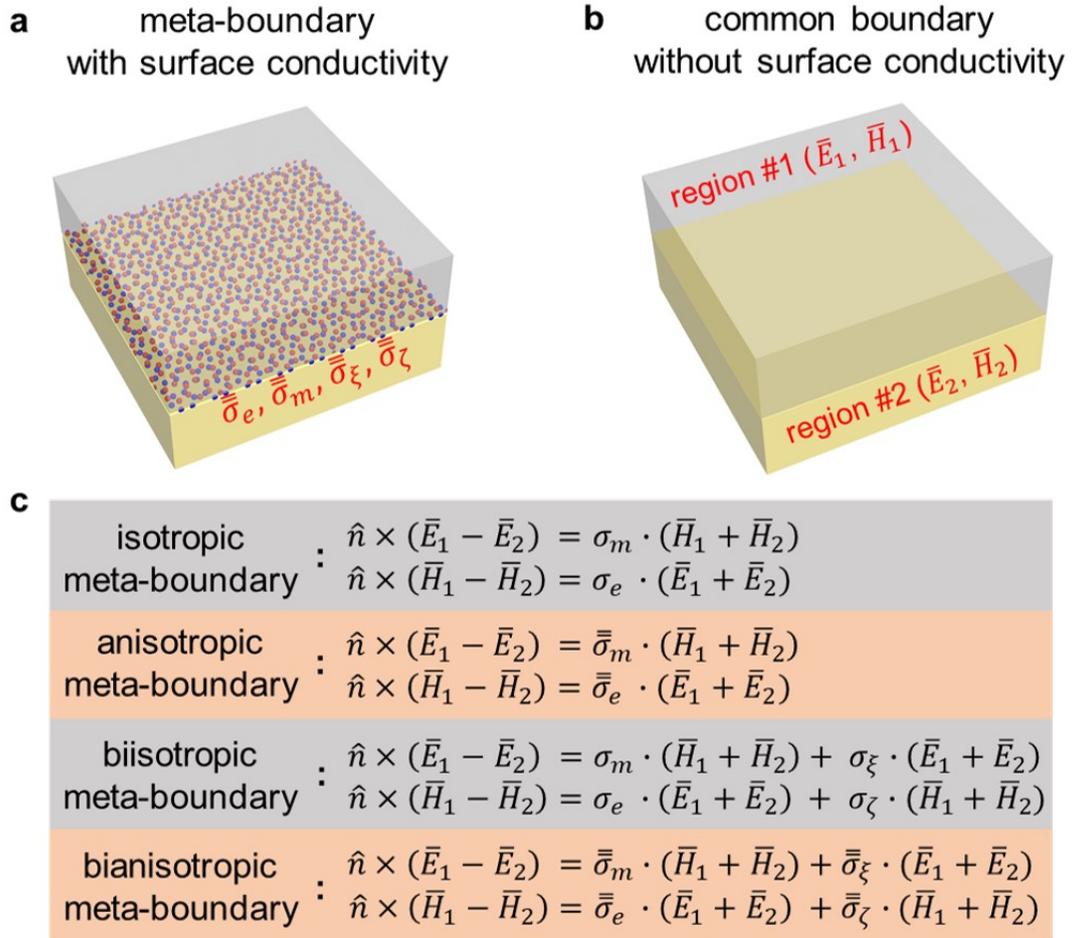

**Figure 1** Meta-boundaries. (a, b) Schematic of meta-boundaries and common boundaries. (c) Different types of meta-boundaries.

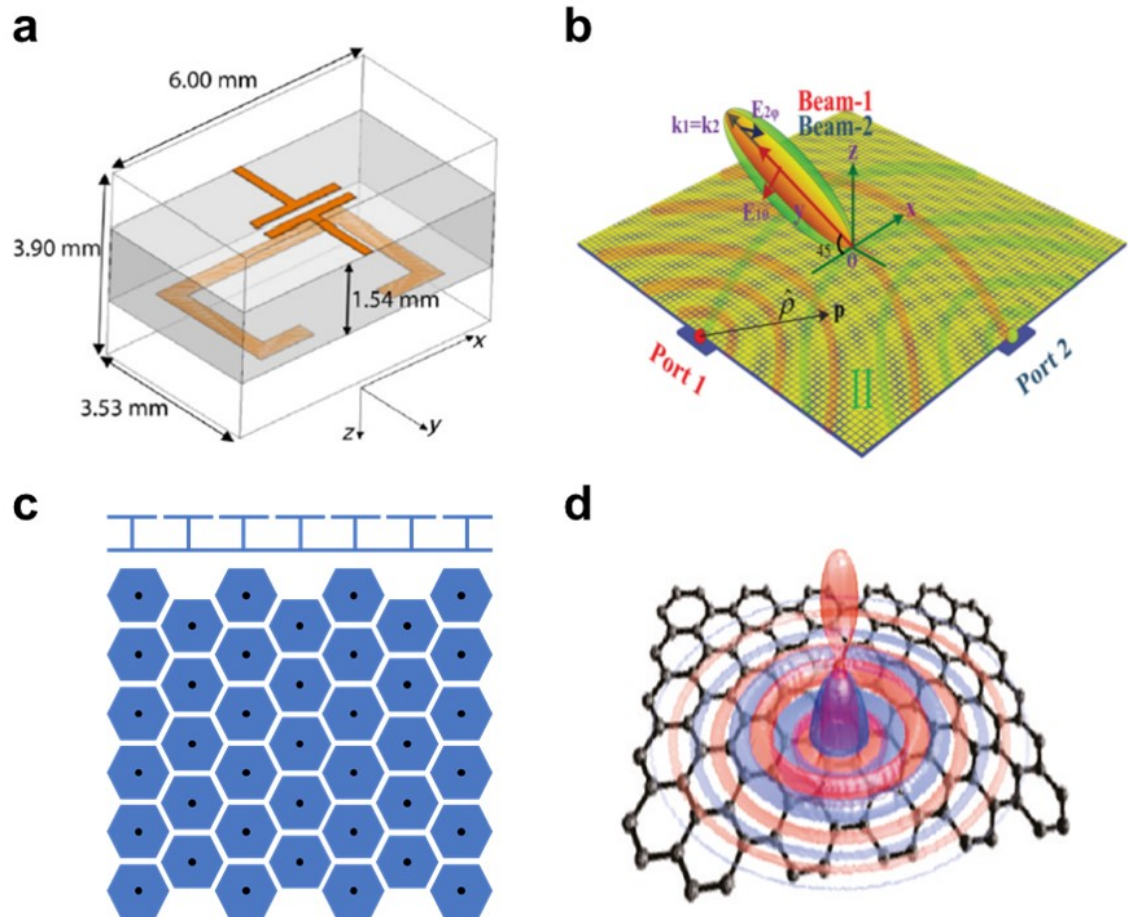

**Figure 2** Isotropic meta-boundaries. (a) Unit cell for Huygens' surface, which can be modelled by an electric surface conductivitity and magnetic conductivitity [80]. (b) Holographic metasurfaces for dual-functional radiations [61]. (c) High impedance metasurface with a mushroom-like unit structure [55]. (d) Near-field excitation of graphene plasmons by depositing a dipole close to the graphene [50].

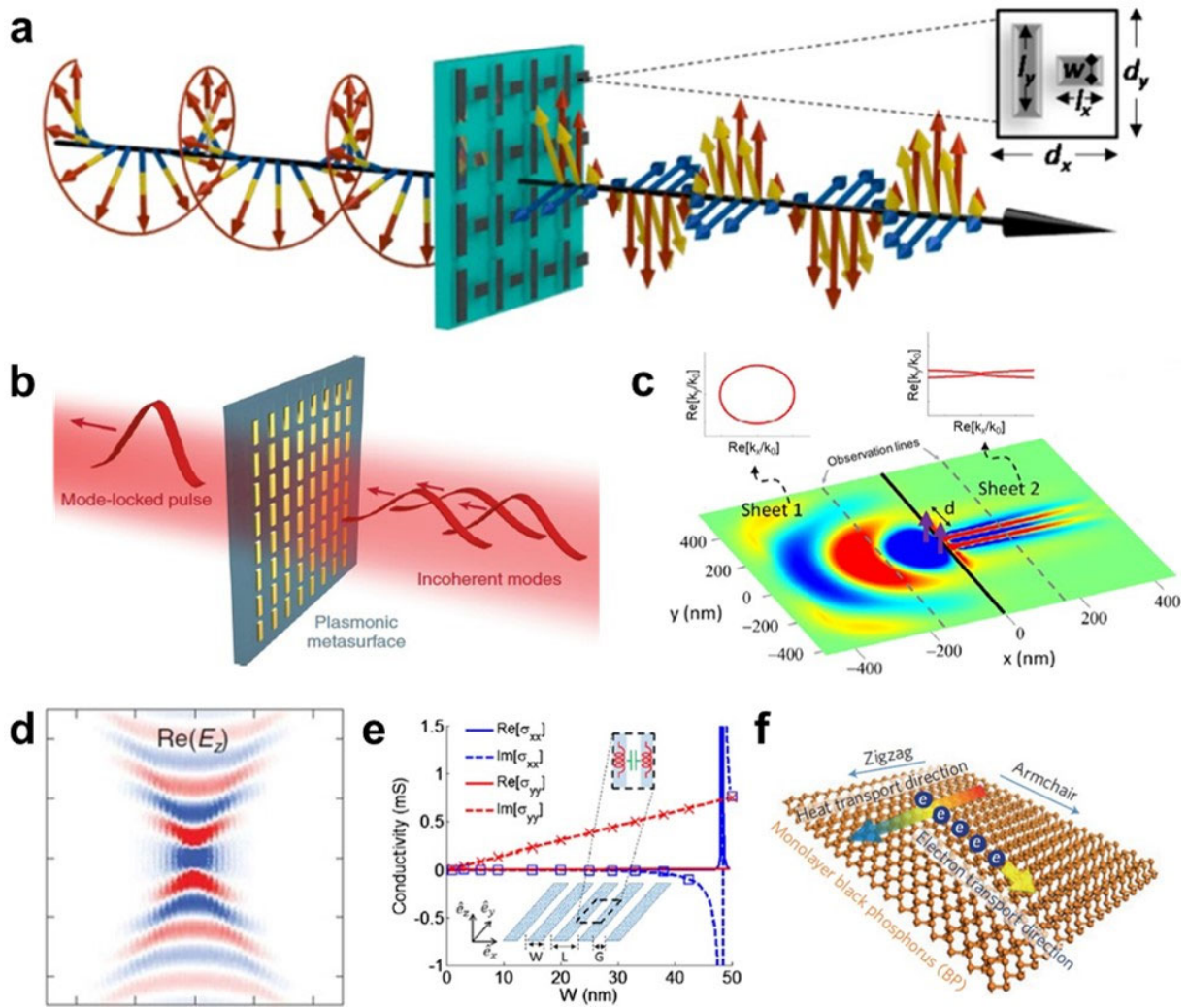

**Figure 3** Anisotropic meta-boundaries. (a) Conversion between the circularly-polarized light and linearly-polarized light by using the anisotropic meta-boundary [103]. (b) Plasmonic metasurface with periodically arranged gold nanorods [104]. (c) Planar hyperlens. The inset shows the isofrequency contours of surface plasmons in the left and right regions [105]. (d) Near-field excitation of surface waves by putting a dipole close to a hyperbolic metasurface [47]. (e) Surface conductivity of hyperbolic metasurfaces constructed by graphene nanoribbons [109]. (f) Monolayer black phosphorus [125].

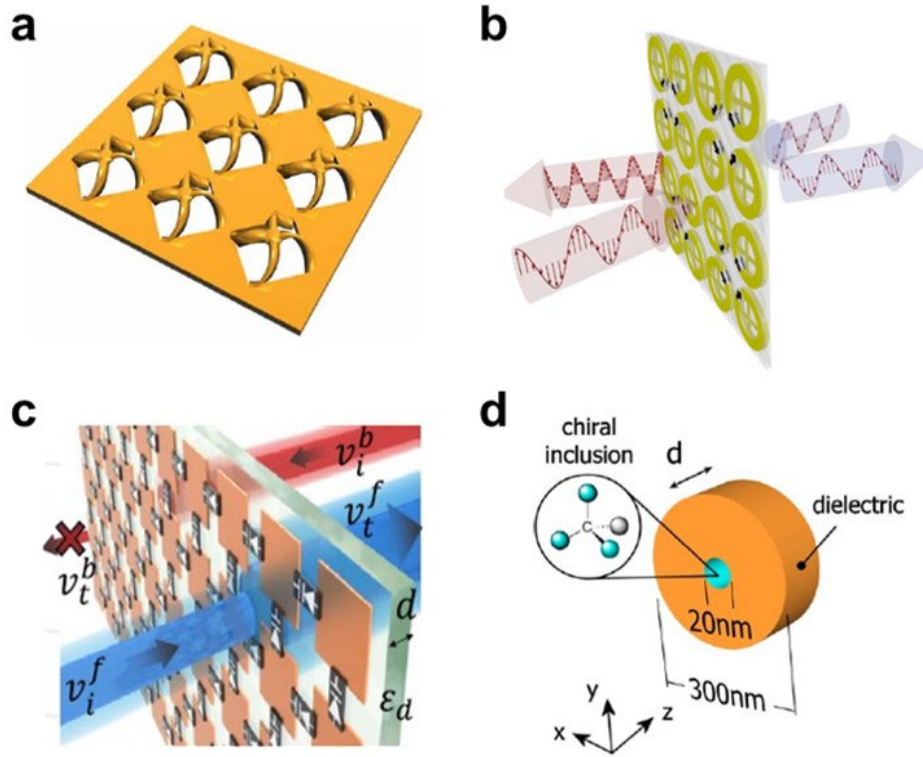

**Figure 4** Biisotropic meta-boundaries. (a) Pinwheel-like metasurface [129]. (b) Bilayer biisotropic metasurface [130]. (c) Time-modulated metasurface comprised of metallic patches array and parallel capacitors [131]. (d) Chiral metasurfaces with their unit cell constructed by a dielectric nanodisk and a chiral inclusion [132].

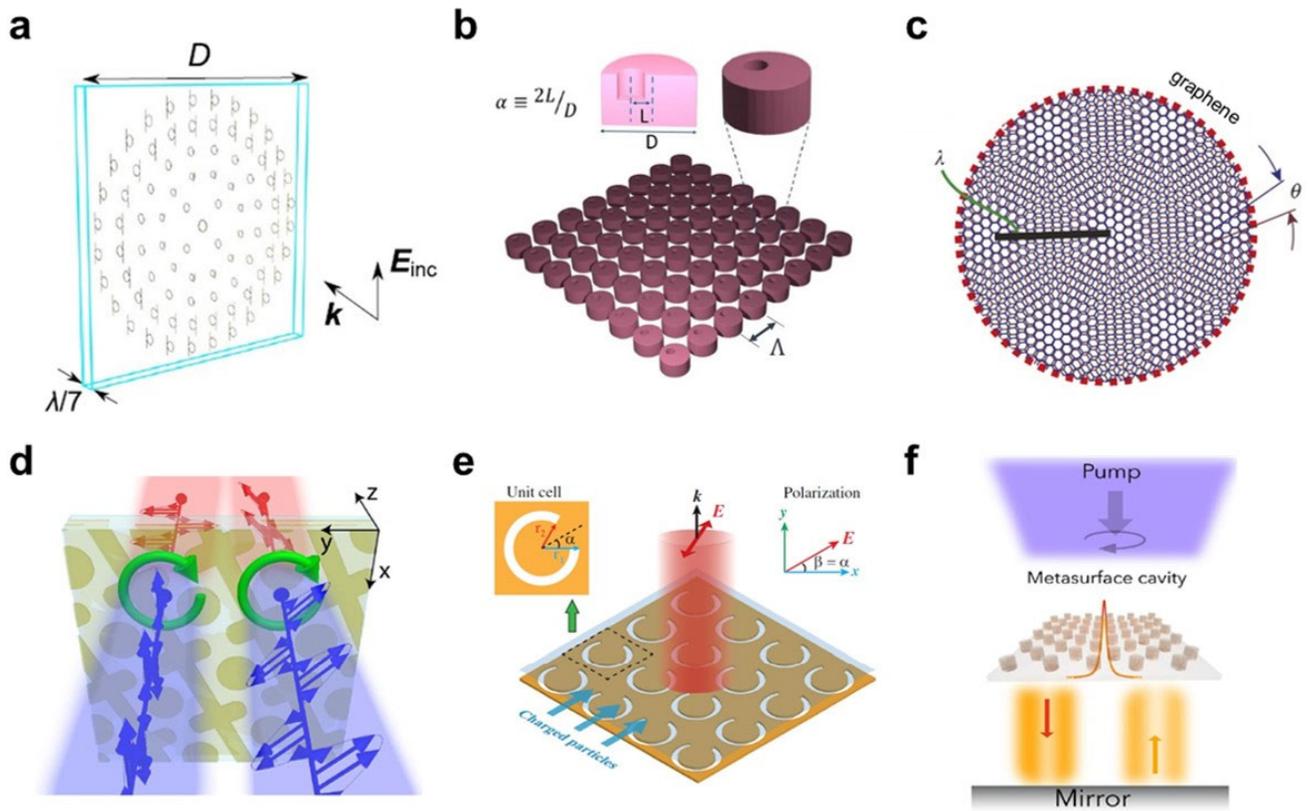

**Figure 5** Bianisotropic meta-boundaries. (a) Functional meta-mirrors using bianisotropic elements [139]. (b) All-dielectric bianisotrapic metasurface [136]. (c) Twisted bilayer graphene [141]. (d) Polarization rotation with ultra-thin bianisotropic metasurfaces [148]. (e) Smith-Purcell radiation from bianisotropic metasurfaces [46]. (f) Self-isolated Raman lasing with a chiral dielectric metasurface [150].

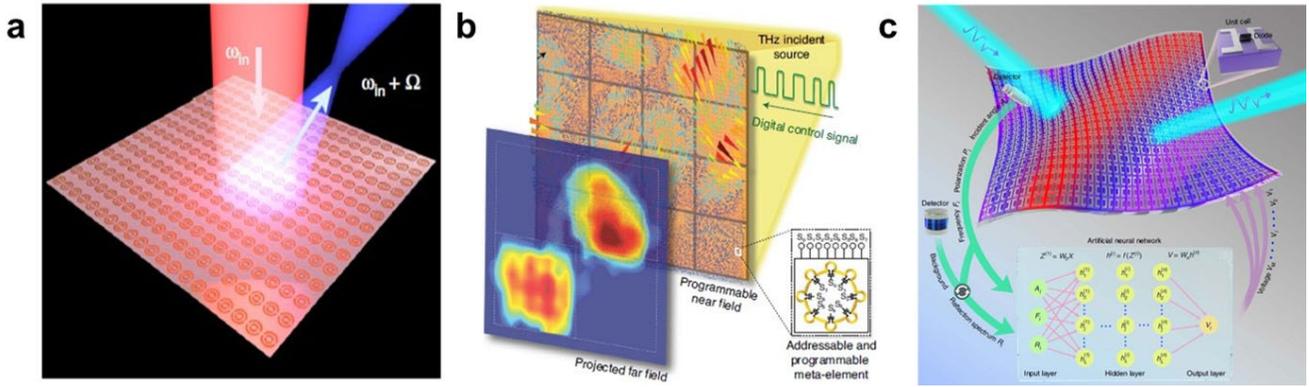

**Figure 6** Active meta-boundaries. (a) Spatiotemporal metasurfaces, which can tune the frequency of reflected light and converge the reflected light at a predesigned focal point [158]. (b) Dynamically programmable metasurfaces, which is made of a programmable two-dimensional array of meta-elements [163]. (c) Self-adaptive metasurface cloak enabled by deep learning [168].

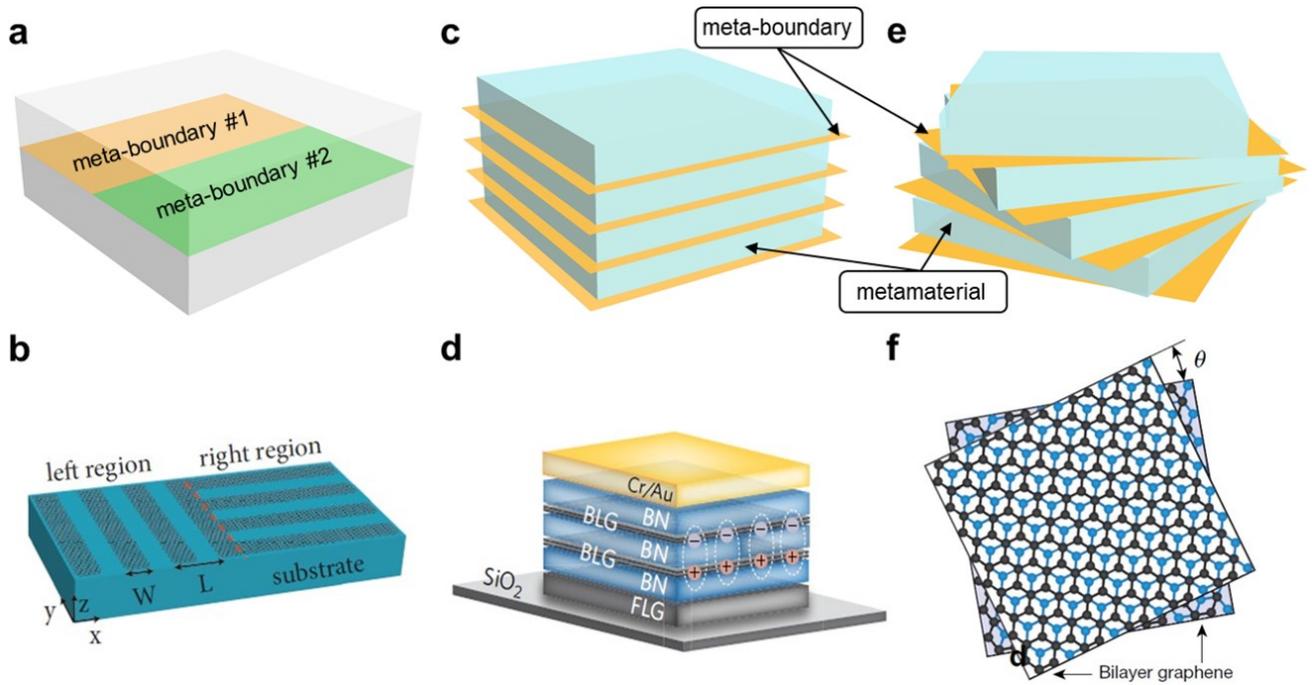

**Figure 7** Multilayer meta-boundaries. (a) Schematic of horizontally stacking multiple meta-boundaries in a same plane. (b) Example of (a), such as the creation of a line boundary by using two hyperbolic metasurfaces with different in-plane rotations [176]. (c) Schematic of vertically stacking meta-boundaries in multiple parallel planes without interlayer twisted angles. (d) Example of (c), such as the construction of multilayer heterostructures by using bilayer graphene and the hyperbolic slab (e.g. hexagonal BN) [177]. (e) Schematic of vertically stacking meta-boundaries in multiple parallel planes with interlayer twisted angles. (f) Example of (e), such as the twisted double bilayer graphene [189].

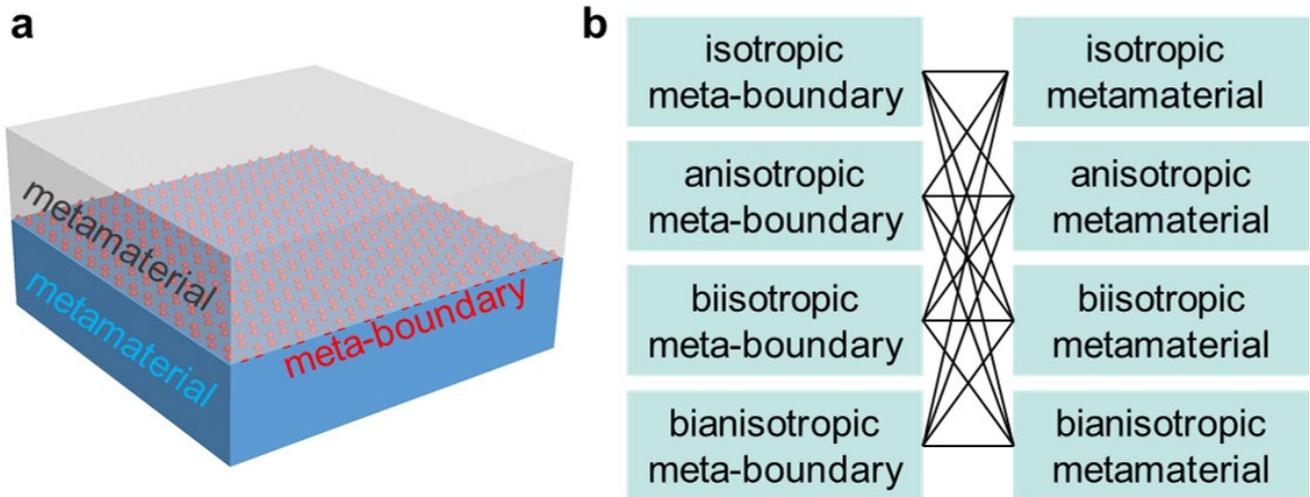

**Figure 8** Controlling light-matter interactions by exploiting meta-boundaries and metamaterials. (a) Structural illustration of a meta-boundary covered and supported by metamaterials. (b) Various combination between meta-boundaries and metamaterials.

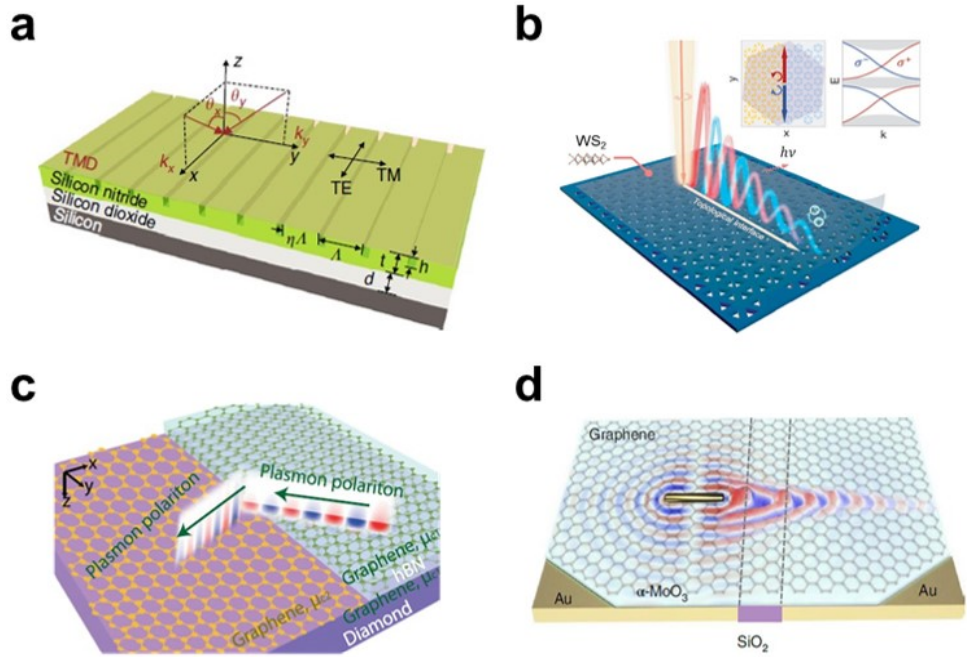

**Figure 9** Examples of controlling light-matter interactions by using meta-boundaries and metamaterials. (a) Generation of exciton-polaritons in monolayer transition metal dichalcogenides (TMD), which is supported by a grating made of silicon nitride [190]. (b) Generation of helical topological exciton-polaritons in tungsten disulfide ($WS_2$), which is supported by a two-dimensional photonic crystal [191]. (c) Negative refraction of highly-squeezed polaritons in a graphene-hexagonal boron nitride (BN) heterostructure [11]. (d) Polaritonic focusing in a graphene/α-phase molybdenum trioxide (α-$MoO_3$) heterostructure [186].